\documentstyle[epsfig,aps,pra]{revtex}
\begin{document}

\newcommand{\ket}[1]{| #1 \rangle}
\newcommand{\bra}[1]{\langle #1 |}
\newcommand{\braket}[2]{\langle #1 | #2 \rangle}
\newcommand{\proj}[1]{| #1\rangle\!\langle #1 |}

\twocolumn[\hsize\textwidth\columnwidth\hsize\csname
@twocolumnfalse\endcsname

]

\centerline {\Large \bf Quantum Computing}
\centerline {\Large \bf with Globally Controlled} 
\centerline {\Large \bf Exchange-type Interactions} 
\smallskip

\centerline {\bf Simon C. Benjamin}
\smallskip
\centerline {Centre for Quantum Computation, Clarendon}
\centerline {Laboratory, University of Oxford, OX1 3PU, UK.}
\smallskip
{\bf
If the interaction between qubits in a quantum computer has a non-diagonal form (e.g. the Heisenberg interaction), then one must be able to ``switch it off'' in order to prevent uncontrolled propagation of states. Therefore, such QC schemes typically demand local control of the interaction strength between each pair of neighboring qubits. Here we demonstrate that this degree of control is not necessary: it suffices to switch the interaction {\em collectively} - something that can in principle be achieved by global fields rather than with local manipulations. This observation may offer a significant simplification for various solid state, optical lattice and NMR implementations.
}
\bigskip

The field of quantum
information processing is presently attracting much interest. Rather than `bits', the fundamental units of classical
information theory, we instead employ `qubits' which represent a general
quantum superposition of `0' and `1'. A computation on a device containing $N$ qubits is a sequence of unitary transformations within its $2^{N}$
dimensional Hilbert space. Quantum
algorithms have been discovered which solve certain problems fundamentally more quickly than any known classical
procedure \cite{SteaneRev}. These algorithms can be described in terms of a sequence of `gates', i.e. elementary operations on single qubits, or between small numbers of qubits. Various sets of gates are known to be universal for quantum computing: a universal set is an efficient set of building blocks for any  algorithm (just as the AND and NOT gates constitute a universal set for classical computation).  Any physical system proposed for quantum computation must be capable of realizing such a set.

The most well know universal set is: the set of all one-qubit gates (i.e. any unitary transformation applied to a single qubit) together with the Control-NOT. The CNOT is the transformation whereby a NOT (i.e. a $\sigma_x$) is applied to one qubit (the `target') if and only if another qubit (the `control') is equal to 1.
Since any any universal set can efficiently implement any other, it follows that any physical implementation must be able to efficiently implement the CNOT. Frequently this is the hardest part of a scheme, since it must involve controlled interaction between two systems.
Indeed, entire QC schemes are often built around a particular idea for controlling such an interaction.
The simplest case is if the interaction is {\em diagonal} in the computational basis (the binary basis formed by each qubit being either in state 0 or in state 1). The Ising interaction, $\sigma_1^z.\sigma_2^z$, is the archetypal example: nuclei in a molecule typically have an (effective) interaction of this form. There is no need to `switch-off' a diagonal interaction, because it does not directly allow states to propagate from one qubit to another. Moreover it has been noted \cite{LloydScience,1DCA} that such a system can be controlled via a cellular-automata (CA) approach whereby it is never necessary to physically address individual qubits (except for the qubit at the `end' of the array).
 
However, diagonal interactions are a special case, and frequently occur in the limit of weak interaction strength.
Approaches which exploit non-diagonal interactions, such as the  the Heisenberg interaction \cite{3qubitExchangeOnly,levy}, typically do so by switching it on and off between specific qubits. In solid-state schemes, this means that there must be at least one independent control electrode between each adjacent pair of qubits \cite{Kane,eSpinQC}. The use of such electrodes has several potential drawbacks: it puts aggressive demands on the fabrication process, it introduces a minimum length scale for qubit separation, and most importantly it provides decoherence channels.
Do we need to have this degree of control when the interactions are non-diagonal? We cannot tolerate having the interaction `always on' between all neighboring qubits, since then our 0's and 1's will uncontrollably `flow' through the system (e.g. as spin-waves). However, perhaps we need not switch off {\em individual} qubit interactions {\em independently}? If it suffices to switch on/off many interactions simultaneously, then with clever qubit design we might achieve this switching with global fields - we could then adopt the CA approach and dispense with local control electrodes \cite{knowItIsPossibleByMapping}.

Here we will give one method of achieving this. The architecture we employ (see Fig 1(a)) is the simplest that is compatible with the basic idea. Our computer is a one-dimensional array of two-state systems, or `cells'. Note that the cells are labeled in the pattern ABAB... (A and B  are not necessarily physically different). We imagine that the cell at one end of the array is `special' - for example, it might be independently controlled and associated with a measuring device - but that for all other qubits the following description applies. 
The Hamiltonian of this system is $H_{tot}=\sum_{i=1}^{n} H^s_i+\sum_{i=1}^{n-1}H^{int}_{i,i+1}$, where the former term represents the energy of each qubit in isolation, and the latter represents the interaction between neighboring qubits. It is well known that if we can freely control the magnitude of these $2n-1$ terms, then we can perform universal QC. However, we now introduce an extreme simplification: $H^s_{2i}$=$H^A$, $H^s_{2i+1}$=$H^B$, $H^{int}_{2i,2i+1}$=$H^{AB}$ and $H^{int}_{2i-1,2i}$=$H^{BA}$. Thus we have gone from $2n-1$ independent terms to just $4$ independent terms: at {\em all times} the single qubit energies of all the A cells are the same, similarly the B cells all have the same single qubit energy, and all the interaction terms have one of two forms, denoted $H^{AB}$ or $H^{BA}$, with the type alternating along the array.

We need not specify these Hamiltonians, they may have any form provided that the following conditions are met:

\noindent {\bf (a)} It must be possible to switch ``off'' $H^{BA}$, so that the system then decouples into a set of (identically interacting) A-B pairs. 

\noindent {\bf (b)} By manipulation of the remaining terms $H^A$, $H^B$ and $H^{AB}$, it must be possible to implement any two-qubit operation on the A-B pairs (all pairs will experience the same manipulation, of course).

\noindent {\bf (c)} As requirements (a) and (b), but `switching off' the $H^{AB}$ interaction instead.

The design challenge will be to meet condition (a) without re-introducing local gates - several possibilities are discussed later. Condition (b) is relatively easy: it could be met, for example, by a pair of systems A and B having different g-factors and being coupled by a Heisenberg interaction. (This is an example where A and B are physically distinct systems). With a short sequence of steps, involving varying the Heisenberg coupling strength and altering the orientation of a global magnetic field, it is possible to perform any desired two-qubit gate on this isolated pair \cite{myPaper}.

These ingredients then suffice to perform universal quantum computation. We will briefly describe how this may be achieved; the procedures are conceptually similar to those discussed in Ref \cite{1DCA} (in fact they are rather more simple). In the following, we will use the term `$\alpha$-phase' to refer to the system when $H^{BA}$=$0$, and `$\beta$-phase' to refer to the $H^{AB}$=$0$ case.
In Fig 1(b), successive rows of cells show how the state of the array changes over time. Consider the top-most row. Here there are two qubits, denoted Y and Z, stored in the illustrated section of the array. All-but-one of the other cells are are in state `0' - the one exception is a single cell is in state 1, which is said to be representing the `control unit' (CU).
Notice that only the $A$ cells are representing qubits, and the CU is  represented by a $B$ cell. Now suppose that we fix $H^{BA}=0$, i.e. adopt the $\alpha$ phase, and perform a SWAP operation between the (now isolated) $AB$ pairs. We denote this operation as $\alpha^{\rm SWAP}$.
The result of this operation is shown in the second row: all the information (qubits) is shifted one cell to the right, and the CU is shifted one cell to the left. If we now perform the complementary operation $\beta^{\rm SWAP}$, then the qubits will shifted one cell {\em further} to the right, and the CU one further to the left, as shown in the third row.  Because we are using pure SWAP operations, when qubit $Z$ `collides' with the CU, the two objects simply pass `through' one another undisturbed. Therefore with an appropriate sequence of $\alpha^{\rm SWAP}$ and $\beta^{\rm SWAP}$ operations, we can move the CU back and forth through the qubits as we wish. Now suppose that we are `running' some quantum algorithm which calls for a transformation $U$ to be applied only to one specific qubit - qubit $Y$, say. This ``single-qubit gate'' operation is shown in detail in Fig 1(b), and schematically in Fig 1(c). We first move the CU until it is adjacent to the qubit. We then perform a control-U between the CU (acting as the control qubit) and Y (the target qubit). We established in our list of requirements that it is possible to do this. Since the CU is in state $\ket{1}$, the transformation $U$ {\em will} be applied to qubit X; all other qubits will of course also be subject to the control-U, however since their controlling qubits will be in state `0' they will not undergo any net transformation. 
Now we are free to move the CU away to its next destination qubit.

\begin{figure}
\centerline{\epsfig{file=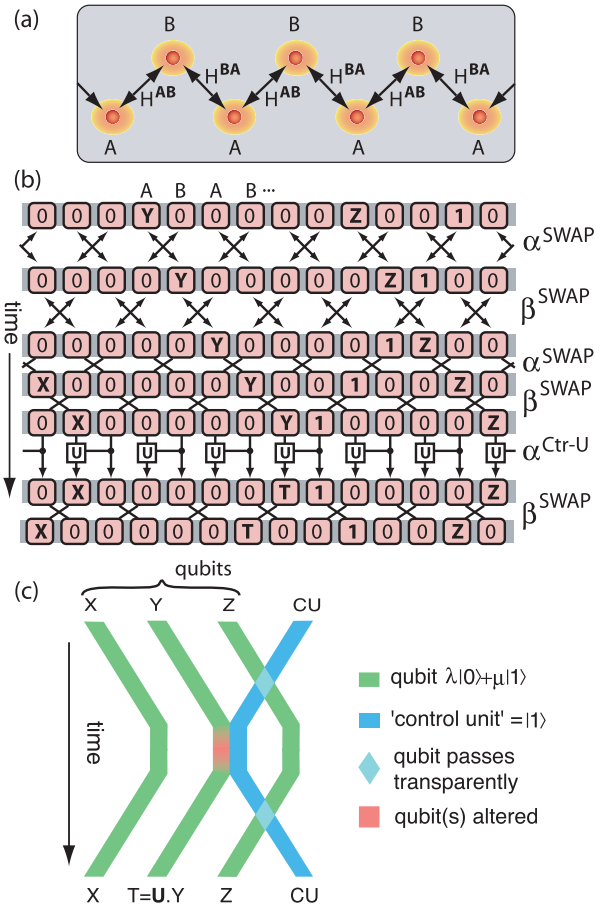,width=8cm}}
\vspace{0.2cm}
\caption{(a) A sketch of the proposed architecture: a one-dimensional array of 2-state cells. All the cell-cell interactions denoted $H^{AB}$ are switched simultaneously, as are all interactions denoted $H^{BA}$. (b) An explicit description of the steps involved in applying a single-qubit `gate' $U$ to  qubit Y. Successive rows show how the state of a section of the array develops over time. Notation on the right indicates the elementary operation applied at each interviewing step: $\alpha$ indicates that only the $H^{AB}$ interactions are non-zero, and similarly $\beta$ indicates only $H^{BA}$ are non-zero. (c) Schematic summarizing the flow of information in (b).}
\label{figure1}
\end{figure}

The process for performing a control-U {\em between two of the qubits} is rather more complex. It is shown explicitly in Figure 2. Essentially, the CU goes first to the control qubit ($Y$ in the example of Fig 2), and once there the CU itself undergoes a transformation: it is switched to zero if, and only if, $Y$=$0$. In other words, $Y$ is copied (not cloned) onto the cell representing the CU. This transformation is chosen so as not cause any net effect for any of the other qubits (therefore a simple CNOT followed by NOT would not suffice: this would have the desired effect on the CU but would also duplicate all other qubits). The transformation has a disruptive effect on $Y$, leaving it displaced and transformed - but this disruption will later be undone. Once the transformation is complete, the new CU is moved to the target qubit, and a control-U is applied as in the case of the single-qubit gate in Fig 1. Of course, if the CU has been switched to zero, i.e. if $Y=0$, then there will be no net effect. Now {\em all previous steps except the last are repeated in reverse order}. In this way the disruption of the Y-qubit is unmade, and the CU is restored. The system has thus returned to its state at the beginning of the procedure, {\em except} that the desired control-U has been performed between the $Y$ and $W$.

\begin{figure}
\centerline{\epsfig{file=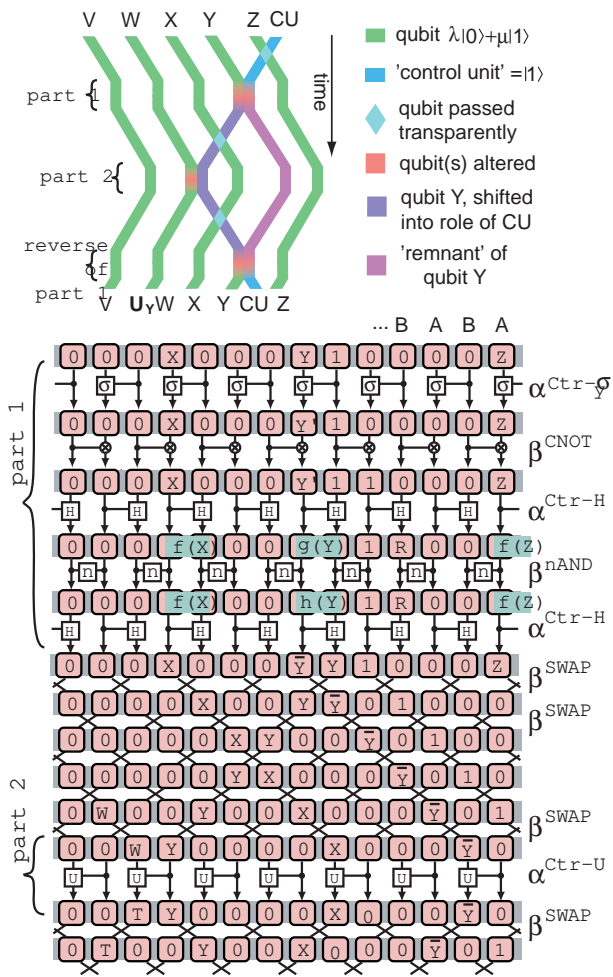,width=8cm}}
\vspace{0.2cm}
\caption{(a) Schematic showing the process needed to perform Control-U between two qubits: here the control qubit is $Y$ and the target is $W$. (b) The explicit depiction, analogous to Fig. 1(b). Here the following additional elementary two-qubit operations are employed:  controlled Hadamard (denoted Ctrl-H), controlled NOT (denoted CNOT), controlled Pauli-Y (denoted Ctrl-$\sigma_y$), and the phase inverting gate (denoted nAND) which acts when both inputs are $1$. 
}
\label{figure2}
\end{figure}

As described so far, the computer is limited to performing gates in series: the CU can only be performing one task at a time. However, the technique of `sub-computations' described in Ref. \cite{1DCA} can be applied here to provide a plurality of CU's within the computer, so that many gates can be performed simultaneously (the CU's are turned on and off as required, all within the fundamental requirement of pure global control). Ref. \cite{1DCA} also discusses the mechanisms for initial state preparation, and for intermediate and final measurements, which we would adopt here.

A significant cost associated with using this scheme lies in the number of zero-state cells required as `padding' between the qubits. In order to to use the specific sequence of steps specified in Fig's 1 \& 2, it is necessary that each qubit have at least $5$ `blank' cells on one side, and at least $3$ on the other. Therefore the most compact way of representing qubits is P(3-blanks)Q(5-blanks)R(3-blanks)S(5-blanks)...., which implies an average of five cells per qubit. This is an appreciable factor, but it might easily be justified by the benefit of doing-away with local gates. Moreover, the sequence labeled `part 1'  in Fig. 2 was found by hand, and may not be optimal - there may be sequences that require less work space. Also note that we have used the simplest architecture that can support the basic idea; more complex architectures would allow a more compact representation of the qubits. Examples of superior architectures would include: a two-dimensional array, an array with more states per cell, or an array 
where there are more than two sets of switchable interactions (e.g. an array with interactions $H^{AB},H^{BC},H^{CA},H^{AB}...)$. There are other possible benefits to the use of a more complex architecture: it might permit the present scheme to be combined with the concept of all-Heisenberg switching \cite{3qubitExchangeOnly,levy} (where no parameters other than the qubit-qubit interaction strength need be varied in order to perform universal computation), and it might similarly enable the related concept of decoherent free subspaces \cite{DecoherenceFree}. 

This concludes the main purpose of this Letter, which was to discuss the exploitation of non-diagonal qubit-qubit interactions {\em without} the use of local control elements. We now conclude with some ideas about how this conceptual architecture might be put into practice. The first candidates one thinks of are the optical lattice schemes \cite{atom1,atom2}. Here there are strong prospects for encoding two subsets of atomic `species', and subsequently controlling their relative position and hence their interaction. Therefore this implementation seems very promising. A second possibly would be to modify one of the solid state schemes that involve gated Heisenberg interactions. These proposals typically use local electrodes to alter the wavefunction overlap between two neighboring electron spins \cite{Kane, eSpinQC}. Figure 3 shows a set of speculative sketches, showing how the use of shaped confinement potentials (e.g., quantum dots) could provide the degree of control required for the present scheme, purely by varying an external $E$ field. As an alternative to using shaped confinement potentials, one could think of an array where the cells are alternately represented by charge carriers of different sign (e.g. electrons and holes), so that an applied $E$ field causes the $A$ and $B$ arrays to move in different directions. In all cases our fundamental requirement is to switch an array from an $AB$ pairing to a $BA$ pairing. There may be many other methods of achieving this, but further speculation is beyond the scope of the present text: these few example should demonstrate that our two-phase architecture is, in principle,  a physically plausible one.

\begin{figure}
\centerline{\epsfig{file=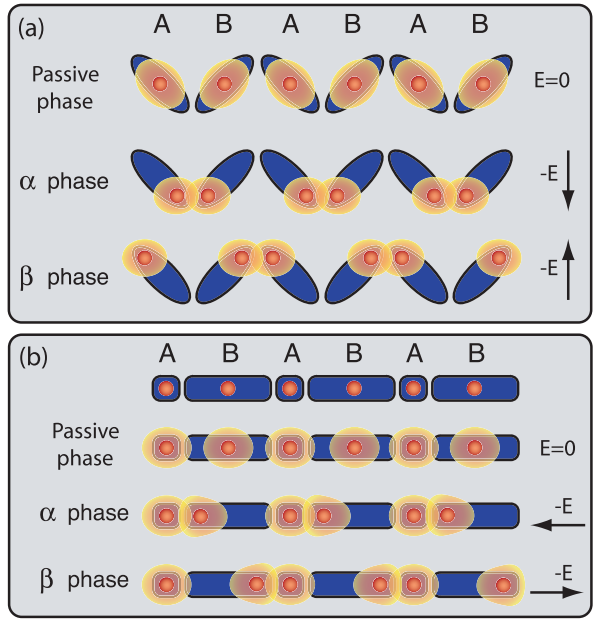,width=7.8cm}}
\vspace{0.2cm}
\caption{Two speculative sketches showing how the use of shaped confinement potentials (blue) might provide the required degree of control over the qubit-qubit interactions. In these figures we may imagine that our two-state system (orange/yellow) is the spin of a single electron. (a) Here the confining potentials (e.g. quantum dots) are  shaped so that there is only significant wavefunction overlap when there is an applied field. Moreover, the sign of the field determines whether the qubits couple in the pattern AB, AB .., or the pattern ..BA, BA..  (b) An alternative based on a sharp confinement potential alternating with an elongated one.}
\label{figure3}
\end{figure}

As a final remark, we note that although we have stressed the possibility of doing away with local control entirely, this is certainly not a {\em requirement} of the present scheme. Indeed, one interesting possibility would be to hybridize the ideas presented here with an electrode-based proposal, in order to simply {\em reduce} the required  density of electrodes.

This work was supported by an EPSRC fellowship.


\begin{references}
\bibitem{knowItIsPossibleByMapping} In fact, we can establish that this must be possible simply by noting that the  the primitive operations used in Ref. \cite{1DCA} (such as ``Cells of type $A$: invert your state iff both your neighbors are in state 1'') can be translated into sequences of primitives of the kind used in Fig's 1 and 2 here. However, this `emulation' of the earlier scheme gives a rather bulky implementation: instead the procedures described in this paper are `optimized' for the strengths of the present architecture.
\end{references}
\end{document}